# *TIME* AS A QUANTUM OBSERVABLE

V.S.Olkhovsky,*  E.Recami,**

(*) *Institute for Nuclear Research of NASU, Kiev-03028, Ukraine.*

(**) *Facoltà di Ingegneria, Università statale di Bergamo, Bergamo, Italy;*
and *INFN-Sezione di Milano, Milan, Italy.*

**Abstract**: Some results are reviewed and developments are presented on the study of *Time* in quantum mechanics as an observable, canonically conjugate to *energy*. Operators for the observable Time are investigated in particle and photon quantum theory. In particular, this paper deals with the *hermitian* (more precisely, maximal hermitian, but non-selfadjoint) operator for Time which appears: (i) for particles, in ordinary non-relativistic quantum mechanics; and (ii) for photons (i.e., in first-quantization quantum electrodynamics).

### 1. Introduction. An operator for Time in quantum physics for non-relativistic particles and for photons.

Almost from the birth of quantum mechanics (see, for example, ref.[1]) it is known that Time cannot be represented by a selfadjoint operator, with the possible exception of special systems (such as an electrically charged particle in an infinite uniform electric field) [1]. This circumstance results to be in contrast with the known fact that time, as well as space, in some cases plays the role just of a parameter, while in some other cases *is* a physical observable, which *ought* to be represented by an operator. The list of papers devoted to the problem of time in quantum mechanics is extremely large (see, for instance, refs.[2-20], and references therein). The same situation had to be faced also in quantum electrodynamics and, more in general, in relativistic quantum field theory (see, for instance, refs.[6,17]).

As to quantum mechanics, the very first articles can be found quoted in refs.[2-8]. A second set of papers on time in quantum physics[9-20] appeared in the nineties, stimulated mainly by the need of a consistent definition for the tunneling time. It is noticeable, and let us stress it right now, that this second set of papers seems however to ignore Naimark's theorem[21], which had previously constituted an important (direct or indirect) basis for the results in refs.[2-7]. Let us recall that Naimark's theorem states[21] that the non-orthogonal spectral decomposition of an hermitian operator *can be approximated* by an orthogonal spectral function (which corresponds to a selfadjoint operator), in a weak convergence, *with any desired accuracy*: We shall come back to such questions in the following.

---

[1]) The fact that time cannot be represented by a selfadjoint operator is known to follow from the semi-boundedness of the continuous energy spectra, which are bounded from below (usually by the value zero). Only for an electrically charged particle in an infinite uniform electric field, and for other very rare special systems, the continuous energy spectrum is not bounded and extends over the whole energy axis from $-\infty$ to $\infty$. It is curious that for systems with continuous energy spectra bounded from above and from below, the time operator is selfadjoint and yields a discrete time spectrum.

Namely, in refs.[2-4] (more details having been added in refs.[5,6]), and, independently, in [7], it has been shown, by recourse to such an important theorem, that, for systems with continuous energy spectra, *time* can be introduced as a quantum-mechanical observable, canonically conjugate to energy. More precisely, the time operator resulted to be hermitian, even if not selfadjoint: as we are going to see.

The main goal of the present paper is precisely to justify the association of *time* with a quantum observable, by exploiting the properties of the hermitian operators in the case of *continuous* energy spectra, and the properties of quasi-selfadjoint operators in the case of *discrete* energy spectra.

Such a goal is conceptually connected with the more general problem of a four-position operator, canonically conjugate to the four-momentum operator for relativistic spin-zero particles: This more general problem will be examined elsewhere, still starting from results contained in refs.[22-24]. Also other *relevant* sectors of quantum mechanics and quantum field theory, including unstable-state decays, will be considered elsewhere.

## 2. On Time as an observable (and on the Time-Energy uncertainty relation) in non-relativistic quantum mechanics, for systems with continuous energy spectra: A sketchy review.

As we were saying, already in the seventies[2-7] it has been shown that, for systems with continuous energy spectra, the following simple operator, canonically conjugate to energy, can be introduced for time:

$$\hat{t} = \begin{cases} t & \text{in the (time) } t\text{-representation,} \quad (1a) \\ -i\hbar \dfrac{\partial}{\partial E} & \text{in the (energy) } E\text{-representation} \quad (1b) \end{cases}$$

which is not selfadjoint, but is *hermitian*, and acts on square-integrable space-time wavepackets in representation (1a), and on their Fourier-transforms in representation (1b), once the point $E=0$ is eliminated (i.e., once one deals only with moving packets, excluding any *non-moving* rear tails and the cases with zero flux). [2]. In refs. [2-6], the operator $\hat{t}$ (in the *t*-representation) had the property that any averages over time, in the one-dimensional (1-D) scalar case, were to be obtained by use of the following *measure* (or weight):

$$W(x,t)\mathrm{d}t = \frac{j(x,t)\mathrm{d}t}{\int_{-\infty}^{\infty} j(x,t)\mathrm{d}t}, \quad (2)$$

where the (*temporal*) probability interpretation of the flux density $j(x,t)$ corresponds to the probability for a particle to pass through point $x$ during the unit time centered at $t$, when travelling in the positive *x*-direction.. Such a measure is not postulated, but is a direct consequence of the well-known probabilistic (*spatial*) interpretation of $\rho(x,t)$, and of the continuity relation $\partial\rho(x,t)/\partial t + \text{div } j(x,t) = 0$. Quantity $\rho(x,t)$ is, as usual, the probability of finding the considered moving particle inside a unit space interval, centered at point *x*, at time *t*.

---

[2] Such a condition is enough for operator (1a,b) to be an *hermitian (*or, more precisely, *"maximal hermitian"*) *operator*[5-7] (see also [16,17]), according to Akhiezer & Glazman's terminology[26]. Let us explicitly notice that, anyway, this physically reasonable boundary condition $E \neq 0$ can be dispensed with, by having recourse to bi-linear operators, as shown by us in Appendix B.

Quantities $\rho(x,t)$ and $j(x,t)$ are related to the wave function $\Psi(x,t)$ by the usual definitions $\rho(x,t)=|\Psi(x,t)|^2$ and $j(x,t) = \text{Re }[\Psi^*(x,t)(\hbar/i\mu)\partial\Psi(x,t)/\partial x]$. When the flux density $j(x,t)$ changes its sign, the quantity $W(x,t)dt$ is no longer positive definite and, as it was known in refs.[16,17], it acquires the physical meaning of a probability density *only* during those partial time-intervals in which the flux density $j(x,t)$ does keep its sign. Therefore, let us introduce the *two* measures[16,17,6], by separating the positive and the negative flux-direction values (i.e., the flux signs):

$$W_\pm(x,t)dt = \frac{j_\pm(x,t)dt}{\int_{-\infty}^{\infty} j_\pm(x,t)dt} \tag{2a}$$

with $j_\pm(x,t)=j(x,t)\theta(\pm j)$.

We would like now to make a useful *generalization* (cf. refs.[2-6,17]) of the definitions of the averages over time $< t^n >$, with $n=1,2,3,\ldots$, for $<f(t)>$, quantity $f(t)$ being *any* arbitrary *analytic* function of time; and write down, by using the weight (2), the single-valued expression

$$<f(t)> = \frac{\int_{-\infty}^{\infty} j(x,t)f(t)dt}{\int_{-\infty}^{\infty} j(x,t)dt} = \frac{\int_0^\infty dE \frac{1}{2}[G^*(x,E)f(\hat{t})vG(x,E) + vG^*(x,E)f(\hat{t})G(x,E)]}{\int_0^\infty dEv|G(x,E)|^2}, \tag{3}$$

in which $G(x,E)$ is the Fourier-transform of the moving 1-D wavepacket

$$\Psi(x,t) = \int_0^\infty G(x,E)\exp(-iEt/\hbar)dE = \int_0^\infty g(E)\varphi(x,E)\exp(-iEt/\hbar)dE \tag{3'}$$

when going on from the time to the energy representation.[3]. For free motion, $G(x,E)=g(E)\exp(ikx)$; $\varphi(x,E)=\exp(ikx)$; and $E=\mu\hbar^2k^2/2=\mu v^2/2$; with the normalization condition

$$\int_0^\infty v|G(x,E)|^2 dE = \int_0^\infty v|g(E)|^2 dE = 1$$

and the boundary conditions

$$\left[\frac{d^n g(E)}{dE^n}\right]_{E=0} = \left[\frac{d^n g(E)}{dE^n}\right]_{E=\infty} = 0, \quad \text{for } n=0,1,2,\ldots \tag{4}$$

Conditions (4) imply a very rapid decrease, till zero, of the flux densities near the boundaries $E=0$ and $E=\infty$ : this complies with the actual conditions of real experiments, and therefore they do not represent any restriction of generality (anyway, see footnote [2]).

---

[3] Let us recall that in this Section we are confining ourselves to systems with continuous spectra only.

In eq.(3), $\hat{t}$ is defined through relation (1b). One should notice that relation (3) expresses the *equivalence* of the *time* and of the *energy representations* (with their own appropriate averaging weights). This equivalence is a consequence of the existence of the time operator. In quantum mechanics, for the time and energy operators it appears to hold the same formalism as for all other pairs of canonically-conjugate observables.

For quasi-monochromatic particles, when $|g(E)|^2 \approx K \delta(E-\overline{E})$, with $K$ a constant, equation (3) gets the simpler expression

$$\langle f(t) \rangle \equiv \frac{\int_{-\infty}^{\infty} j(x,t) f(t) \mathrm{d}t}{\int_{-\infty}^{\infty} j(x,t) \mathrm{d}t} \approx \frac{\int_{-\infty}^{\infty} \rho(x,t) f(t) \mathrm{d}t}{\int_{-\infty}^{\infty} \rho(x,t) \mathrm{d}t} = \frac{\int_{0}^{\infty} \mathrm{d}E\, G^*(x,E) f(\hat{t}) G(x,E)}{\int_{0}^{\infty} \mathrm{d}E\, |G(x,E)|^2} \qquad (3a)$$

because of the relations $j(x,t) \approx v\rho(x,t) \approx \overline{v}\rho(x,t)$.

The two canonically conjugate operators, the time operator (1) and the energy operator [4]

$$\hat{E} = \begin{cases} E & \text{in the energy } (E\text{-}) \text{ representation,} \\ i\hbar \dfrac{\partial}{\partial t} & \text{in the time } (t\text{-}) \text{ representation,} \end{cases} \qquad (5)$$

satisfy the typical commutation relation[2-7,17]

$$[\hat{E}, \hat{t}] = i\hbar. \qquad (6)$$

Although up to now the Stone and von Neumann theorem[25] has been interpreted as establishing that expressions like (1) and (5) hold for selfadjoint canonically conjugate operators only, actually that theorem is applicable to hermitian operators too: as it has been shown, e.g., in refs.[2-7,17] by utilizing the peculiar mathematical properties of the "maximal hermitian" operators, described in detail in refs.[21,22,26]. Indeed, from eq.(6) the uncertainty relation

$$\Delta E\, \Delta t \geq \hbar/2 \qquad (7)$$

(where the standard deviations are $\Delta a = \sqrt{Da}$, quantity $Da$ being the variance $Da = \langle a^2 \rangle - \langle a \rangle^2$; and $a \equiv E, t$, while $\langle \ldots \rangle$ denotes the average over $t$ with the measures $W(x,t)\mathrm{d}t$ or $W_\pm(x,t)\mathrm{d}t$ in the $t$-representation[3]) was derived also for hermitian operators by a straightforward generalization of the similar procedures which are standard in the case of selfadjoint (canonically conjugate) quantities: see refs [2,4-7].

Moreover, relation (6) satisfies the Dirac "correspondence principle", since the classical Poisson brackets $\{q_0, p_0\}$, with $q_0 = t$ and $p_0 = -E$, are equal to unity[27]. In ref.[5] (see also [6]) it was shown, as well, that *the differences* between the mean times at which a wave-packet passes through *a pair* of points obey the Ehrenfest correspondence principle. Namely, the Ehrenfest

---

[4] The averages over $E$, in the $E$-representation, were performed in refs.[2-7].

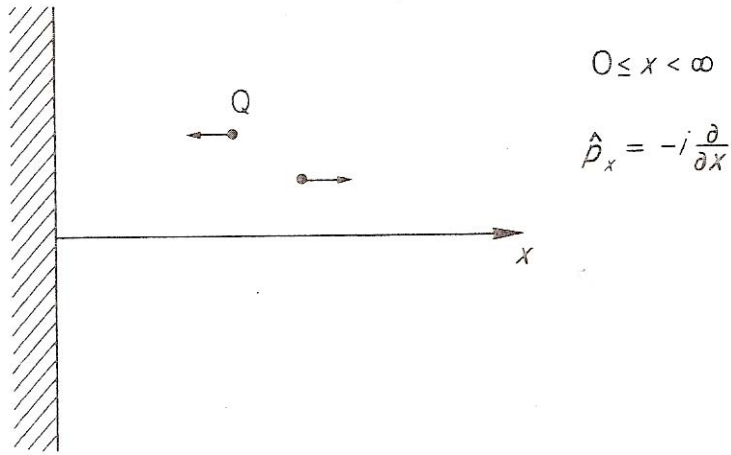

Fig.1 – For a particle Q free to move in a semi-space, bounded by a rigid wall, the operator -i∂/∂x has the clear physical meaning of the particle impulse *x*-component even if it is *not* selfadjoint (cf. von Neumann, [28]).

theorem has been suitably generalized in refs.[5,6].

As a consequence, one can state that, for systems with continuous energy spectra, the mathematical properties[26] of hermitian operators, like $\hat{t}$ in eq.(1), *are sufficient* for considering them as quantum observables: Namely, the *uniqueness* of the "spectral decomposition", also called spectral function (although such an expansion is non-orthogonal), for operators $\hat{t}$, as well as for $\hat{t}^n$ (*n*>1), guarantees the *equivalence* of the mean values of any analytic functions of time evaluated in the *t*- or in the *E*-representation. In other words, such an expansion is equivalent to a *completeness relation* for the (formal) eigenfunctions of $\hat{t}^n$ (*n*>1), which *with any accuracy* can be regarded as orthogonal and correspond to the actual eigenvalues of the continuous spectrum: and these approximate eigenfunctions belong to the space of the square-integrable functions of energy *E*, with the boundary conditions (4) (see the first one of refs.[6], and refs. therein).

From this point of view, there is no *practical* difference between selfadjoint and hermitian (or, more precisely, "maximal hermitian"[26]) operators for systems with continuous energy spectra. Let us repeat that the *mathematical* properties of $\hat{t}^n$ (*n*>1) are quite enough for considering time as a quantum-mechanical observable (like energy, momentum, space coordinates, etc.) *without having to introduce any new physical postulates*.

It is remarkable that von Neumann himself[28], before confining himself for simplicity to selfajoint operators, stressed that operators like our time $\hat{t}$ may represent physical observables, even if they aren't selfadjoint. Namely, he explicitely considered the example of the operator –*iℏ* (∂/∂x) associated with a particle living in the right-semispace bounded by a rigid wall located at *x*=0; that operator is not selfadjoint (it being applicable to wavepackets defined only on the positive *x*-axis), but nevertheless it obviously corresponds to the *x*-component of the observable *momentum* for that particle. See Fig.1, and AppendixA, where the whole question is more clearly exploited and presented.

Finally, let us go back to the fact, mentioned in footnote [2], that our previously assumed boundary condition $E \neq 0$ can be dispensed with, by having recourse[3,29] to the bi-linear operator

$$\hat{t} = (-i\hbar/2) \frac{\overleftrightarrow{\partial}}{\partial E} \quad \quad (1c)$$

where now $(f, \hat{t} g) \equiv (f, (-i\hbar/2)\frac{\partial}{\partial E} g) + ((-i\hbar/2)\frac{\partial}{\partial E} f, g)$. By adopting expression (1c) for the time operator, the algebraic sum of the two terms in the r.h.s. of the last relation, as well as of $(f, \hat{t} f)$ and in $\int_{-\infty}^{\infty} t\, j(x,t)\mathrm{d}t$, results to be automatically zero at point $E=0$. This question will be briefly discussed in Appendix B.

### 3. On the momentum representation of the Time operator.

Instead of the energy-representation, with $0 < E < \infty$, in eqs.(1)-(4) one can use the momentum $k$-representation (see also ref.[7]), with the advantage that, in the continuous spectrum case, $k$ is not bounded, $-\infty < k < \infty$, and the wavepacket writes

$$\Psi(x,t) = \int_{-\infty}^{\infty} \mathrm{d}k\, g(k)\, \varphi(x,k)\, \exp(-iEt/\hbar) , \qquad (8)$$

with $E = \hbar^2 k^2/2\mu$, and $k \neq 0$. In such a case, the time operator (1) (acting on $L^2$-functions of momentum), defined over the whole axis $-\infty < k < \infty$, results to be actually *selfadjoint*, with the boundary conditions

$$\left[\frac{\mathrm{d}^n g(k)}{\mathrm{d}k^n}\right]_{k=-\infty} = \left[\frac{\mathrm{d}^n g(k)}{\mathrm{d}k^n}\right]_{k=\infty} = 0, \qquad n=0,1,2,... , \qquad (9)$$

except for the fact that we have one more to exclude the point $k=0$: an exclusion that we know to be physically and mathematically inessential.

Let us now compare choice (8) with choice (3'). Let us first rewrite eq.(8) as follows:

$$\Psi(x,t) = \int_0^{\infty} \mathrm{d}E\, (E)^{-1/2}\, g((2\mu E)^{1/2}/\hbar)\, \varphi(x, (2\mu E)^{1/2}/\hbar)\, \exp(-iEt/\hbar) +$$
$$+ \int_0^{\infty} \mathrm{d}E\, (E)^{-1/2}\, g(-(2\mu E)^{1/2}/\hbar)\, \varphi(x, -(2\mu E)^{1/2}/\hbar)\, \exp(-iEt/\hbar) . \qquad (10)$$

If we now introduce the two-dimensional weight

$$\widetilde{g}(E) = (\mu/2E\hbar^2)^{1/4} \begin{bmatrix} g((2\mu E)^{1/2}/\hbar) \\ g(-(2\mu E)^{1/2}/\hbar) \end{bmatrix} \qquad (11)$$

then

$$\int_{-\infty}^{\infty} |\Psi(x,t)|^2 \mathrm{d}x = \int_0^{\infty} \mathrm{d}E\, |\widetilde{g}(E)|^2 < \infty \qquad (12)$$

the norm being

$$|\widetilde{g}(E)|^2 = g^*(E) \cdot g(E) > 0 .$$

If wavepacket (8) is travelling only in one direction, that is, $g(k) \equiv g(k)\vartheta(k)$, then the integral $\int_{-\infty}^{\infty} dk$ transforms into the integral $\int_{0}^{\infty} dk$, and the two-dimensional vector goes on to a scalar quantity. In such a case, the boundary conditions (4) can be replaced by relations of the same form, provided that the replacement $E \to k$ is performed.

### 4. An alternative weight for time averages (in the case of a particle *dwelling* inside a certain spatial region).

Let us recall that the weight (2) [as well as its modifications (2a)] has the meaning of a probability for the considered particle to pass through point $x$ during the time interval $(t, t+dt)$. Following the procedure presented in refs.[16,17 and 6] (and refs therein) for the analysis of the equality

$$\int_{-\infty}^{\infty} j(x,t)\, dt = \int_{-\infty}^{\infty} |\Psi(x,t)|^2\, dx, \qquad (13)$$

which evidently follows from the (one-dimensional) continuity relation, one can easily see that an alternative, second weight

$$dP(x,t) \equiv Z(x,t)dx = \frac{|\Psi(x,t)|^2\, dx}{\int_{-\infty}^{\infty}|\Psi(x,t)|^2\, dx} \qquad (14)$$

can be adopted, possessing the meaning of probability for the particle to be "localized" (or to sojourn, i.e., to *dwell*) inside the spatial region $(x,x+dx)$ at the instant $t$, independently from its motion properties. As a consequence, the quantity

$$P(x_1,x_2,t) = \frac{\int_{x_1}^{x_2} |\Psi(x,t)|^2\, dx}{\int_{-\infty}^{\infty}|\Psi(x,t)|^2\, dx} \qquad (14a)$$

will have the meaning of probability for the particle to dwell inside the spatial interval $(x_1, x_2)$ at time $t$.

As it is known (see, for instance, ref.[17] and refs therein), the *mean dwell time* can be written in the *two* equivalent forms:

$$<\tau(x_i, x_f)> = \frac{\int_{-\infty}^{\infty} dt \int_{x_i}^{x_f} |\Psi(x,t)|^2\, dx}{\int_{-\infty}^{\infty} j_{in}(x_i,t)\, dt} \qquad (15a)$$

and

$$<\tau(x_i, x_f)> = \frac{[\int_{-\infty}^{\infty} j(x_f,t)t\, dt - \int_{-\infty}^{\infty} j(x_i,t)t\, dt]}{\int_{-\infty}^{\infty} j_{in}(x_i,t)\, dt}, \qquad (15b)$$

where it has been taken account, in particular, of relation (13), which follows –as we have already seen– from the continuity equation.

Thus, in correspondence with the two measures (2) and (14), when integrating over time, we get *two* different kinds of time distributions (mean values, variances, etc.), possessing *different* physical meanings (which refer to the particle traversal time in the case of measure (2, 2a), and to the particle dwelling in the case of measure (14)). Some examples for 1-D tunneling have been put forth in refs.[16,17].

**5. Extension of the notion of Time as a quantum observable for the case of Photons.**

As it is known (see, for instance, refs.[30], and also ref.[31]), in first quantization the single-photon wave function can be probabilistically described by the wavepacket, in the 1-D case,[5]

$$\vec{A}(\vec{r},t)= \int_{k_0} \frac{d^3 k}{k_0} \vec{\chi}(\vec{k}) \varphi(\vec{k},\vec{r}) \exp(-ik_0 t), \qquad (16)$$

where, as usual, $\vec{A}(\vec{r},t)$ is the electromagnetic vector potential, while $\vec{r}=\{x,y,z\}$; $\vec{k}=\{k_x,k_y,k_z\}$; $k_0 \equiv \omega/c = \varepsilon/\hbar c$; and $k \equiv |\vec{k}|=k_0$. The axis $x$ has been chosen as the propagation direction. Let us notice that $\vec{\chi}(\vec{k}) = \sum_{i=y}^{z} \chi_i(\vec{k})\vec{e}_i(\vec{k})$; with $\vec{e}_i \vec{e}_j = \delta_{ij}$; $x_i, x_j \equiv y,z$; while $\chi_i(\vec{k})$ is the probability amplitude for the photon to have momentum $\vec{k}$ and polarization $\vec{e}_j$ along $x_j$. Moreover, it is $\varphi(\vec{k},\vec{r})=\exp(ik_x x)$ in the case of plane waves; while $\varphi(\vec{k},\vec{r})$ is a linear combination of evanescent (decreasing) and anti-evanescent (increasing) waves in the case of "photon barriers" (i.e., band-gap filters, or even undersized segments of waveguides for microwaves, or frustrated total-internal-reflection regions for light, and so on). Although it is not easy to localize a photon in the direction of its polarization[30], nevertheless for 1-D propagations it is possible to use the space-time probabilistic interpretation of eq.(16), and define the quantity

$$\rho_{em}(x,t)dx = \frac{S_o dx}{\int S_o dx}, \qquad S_0 \equiv \int \int s_0(x,y,z,t)\,dy dz \qquad (17)$$

(quantity $s_o = [\vec{E}^*\vec{E} + \vec{H}^*\vec{H}]/4\pi$ being the energy density, while the electromagnetic field is $\vec{H} = \text{rot}\vec{A}$, and $\vec{E} = -(1/c)\partial \vec{A}/\partial t$) as the *probability density of a photon to be found (localized) in the spatial interval* $(x, x+dx)$ *along axis x at the instant t*; and the quantity

$$j_{em}(x,t)dt = \frac{S_x(x,t)dt}{\int S_x(x,t)dt}, \qquad S_x(x,t) \equiv \iint s_x(x,y,z,t)dydz \qquad (18)$$

(quantity $s_x = c\,\text{Re}[\vec{E}^*\vec{H}]_x / 8\pi$ being the energy flux density) as the *flux probability density of a photon to pass through the point x during the time interval* $(t, t+dt)$: in full analogy with the case of the probabilistic quantities for non-relativistic particles. The justification and convenience of such definitions is self-evident when the wavepacket group-velocity coincides with the velocity of the energy transport. For more general cases, see- ref.[31]. In particular: (i) the wavepacket (16) is quite similar to a wavepacket for non-relativistic particles, and (ii) in analogy with conventional non-relativistic quantum mechanics, one can define the "mean time-instant", for a photon (i.e., an lectromagnetic wavepacket) to pass through point $x$, as follows:

$$<t(x)> = \int_{-\infty}^{\infty} t\, J_{\text{em},x}\, dt = \frac{\int_{-\infty}^{\infty} t S_x(x,t) dt}{\int_{-\infty}^{\infty} S_x(x,t) dt} \ . \qquad (19)$$

---
[5] The gauge condition $\text{div}\vec{A} = 0$ is assumed.

As a consequence [in the same way as in the case of eqs.(1)-(2)], *the form* (1b) *for the time operator* in the energy representation, $-i\hbar\, \partial/\partial E$, *is valid also for photons*, with the same boundary conditions adopted in the case of particles, i.e., with $\chi_i(0) = \chi_i(\infty)$ and with $E = \hbar c k_x$.

The energy density $s_0$ and energy flux-density $s_x$ satisfy the relevant continuity equation

$$\partial s_0 / \partial t + \partial s_x / \partial x = 0$$

which is *Lorentz-invariant* for 1-D spatial propagation [17,31]. It appears evident, therefore, that, even in the case of photons, one can use the same energy representation of the (maximal hermitian) time operator as for particles in nonrelativistic quantum mechanics. So that one can verify the *equivalence* of the expressions for the time standard deviations, the variances, and the mean time durations [evaluated with the measure (18) in the case of photons processes (propagations, collisions, reflections, tunnelings, etc.)] in *both* the time *and* in the energy representations[17, 31]. It is also possible to introduce for photons a second (dwell-time) measure, by extending the procedure exploited for particles in eqs.(14,14a). In other words, *in the cases of* 1-D *photon propagations, time does result to be a quantum observable even for photons*.

**6. Introducing the analogue of the "Hamiltonian" for the case of the Time operator: A new hamiltonian approach.**

In non-relativistic quantum mechanics, the energy operator acquires[6] *two* forms: (i) $i\hbar\, \partial/\partial t$, in the time-representation, and (ii) $\hat{H}(\hat{p}_x, \hat{x},...)$, in the hamiltonian form. The *duality* of these two forms can be easily seen from the Schroedinger equation: $\hat{H}\Psi = i\hbar \dfrac{\partial \Psi}{\partial t}$. One can introduce in quantum theory a similar duality for the case of *time*: Besides the general form (1b) for the Time operator in the energy representation, which is valid for any physical system (in the region of continuous energy spectrum ), one can *express the time operator also in a hamiltonian form*: i.e., in terms of the coordinate and momentum operators, by having recourse to their commutation relations (and by following ref.[32]). Thus, by the replacements

$$\begin{cases} \hat{E} \to \hat{H}(\hat{p}_x, \hat{x},...) \\ \hat{t} \to \hat{T}(\hat{p}_x, \hat{x},...) \end{cases} \qquad (20)$$

and on using the commutation relation (which is similar to eq.(6))

$$[\hat{H},\hat{T}] = i\hbar , \qquad (21)$$

one can obtain, given a specific Hamiltonian, the corresponding explicit expression for $\hat{T}(\hat{p}_x,\hat{x},...)$.

Indeed, this procedure, can be adopted for any physical system with a known Hamiltonian $\hat{H}(\hat{p}_x,\hat{x},...)$, as we are going to see in a concrete example. By going on from the coordinate to the momentum representation, one realizes that the *formal* expressions of *both* the Hamiltonian-type operators $\hat{H}(\hat{p}_x,\hat{x},...)$ and $\hat{T}(\hat{p}_x,\hat{x},...)$ do not change, except for a change of sign in the case of operator $\hat{T}(\hat{p}_x,\hat{x},...)$.

Let us consider, as an explicit example, the simple case of a free particle whose Hamiltonian is

$$\hat{H} = \begin{cases} \hat{p}_x^{\,2}/2\mu, \text{ where } \hat{p}_x = -i\hbar\dfrac{\partial}{\partial x}, & \text{in the coordinate representation,} \quad (22a) \\ p_x^{\,2}/2\mu & \text{in the momentum representation} \quad (22b) \end{cases}$$

whilst, correspondingly, the Hamiltonian-type time operator, in its symmetrized form, writes

$$\hat{T} = \begin{cases} \dfrac{\mu}{2}[\hat{p}_x^{\,-1}x + x\hat{p}_x^{\,-1}], \text{ where } \hat{p}_x^{\,-1} = \dfrac{i}{\hbar}\int dx..., & \text{in the co-ordinate representation,} \quad (23a) \\ \dfrac{-\mu}{2}[p_x^{\,-1}\hat{x} + \hat{x}p_x^{\,-1}], \text{ where } \hat{x} = i\hbar\dfrac{\partial}{\partial p_x}, & \text{in the momentum representation.} \quad (23b) \end{cases}$$

Incidentally, operator (23b) is equivalent to $-i\hbar\,\partial/\partial E$, since $E=p_x^2/2\mu$ ; and therefore it is a (maximal) *hermitian* operator too. Indeed, e.g. for a plane-wave of the type exp*(ikx)* , by applying the operator $\hat{T}(\hat{p}_x,\hat{x},...)$ we obtain the same result in both the coordinate and the momentum representation:

$$\hat{T}\exp(ikx) = \dfrac{x}{v}\exp(ikx), \qquad (24)$$

quantity *x/v* being the free-motion time (for a particle with velocity *v)* for travelling the distance *x*.

**7. Time as an observable (and the Time-Energy uncertainty relation), for quantum-mechanical systems with *discrete* energy spectra.**

Following refs [6,17], for describing the time evolution of non-relativistic quantum systems endowed with a purely *discrete* (or a continuous *and* discrete) spectrum, let us now introduce wavepackets of the form

$$\psi(x,t) = \sum_{n=0}^{\cdots} g_n \varphi_n(x)\exp[-i(E_n-E_0)t/\hbar]  ,  \quad (25)$$

where $\varphi_n(x)$ are orthogonal and normalized bound states; they satisfy the equation $\hat{H}\varphi_n(x) = E_n \varphi_n(x)$, quantity $\hat{H}$ being the Hamiltonian of the system, as well as the condition $\sum_{n=0}^{\cdots} |g_n|^2 = 1$. We omitted a non-significant phase factor $\exp(-iE_0 t/\hbar)$, since it appears in all terms of the sum $\sum_{n=0}^{\cdots}$.

Let us confine ourselves only to the discrete part of the spectrum. Without limiting the generality, we choose $t=0$ as the initial time instant.

Let us firstly consider the simple case of those systems whose energy levels are spaced by intervals which are *multiples* of a "maximum common divisor" $D$. Important examples of such systems are the *harmonic oscillator,* a *particle in a rigid box,* and the *spherical spinning top.* For those systems the wavepacket (25) is a periodic function of time with period $T = 2\pi\hbar/D$ ("Poincaré cycle time"). In the *t*-representation, the relevant energy operator $\hat{H}$ (the Hamiltonian) is a selfadjoint operator acting on the space of the functions $\psi(x,t)$ *periodic* in time, whereas the functions $t\psi(x,t)$, which aren't periodic, do not belong to the same space. On the contrary, in the periodic function space, the Time operator $\hat{t}$ must be itself a periodic function of time $t$, even in the time-representation. This situation is quite similar to the case of the angle, canonically conjugate to angular momentum (see, for instance, refs.[33,34]). Actually, in analogy with the example and results found in ref.[33] for the observable *angle* (possessing a period $2\pi$), let us choose, instead of time $t$, a *periodic* function of time $t$ (possessing as period the Poincaré cycle time $T = 2\pi\hbar/D$ ):

$$\hat{t} = t - T\sum_{n=o}^{\infty} \Theta(t-[2n+1]T/2) + T\sum_{n=0}^{\infty}\Theta(-t-[2n+1]T/2) \quad (26)$$

which is the so-called saw-function of $t$ (see Fig.2).

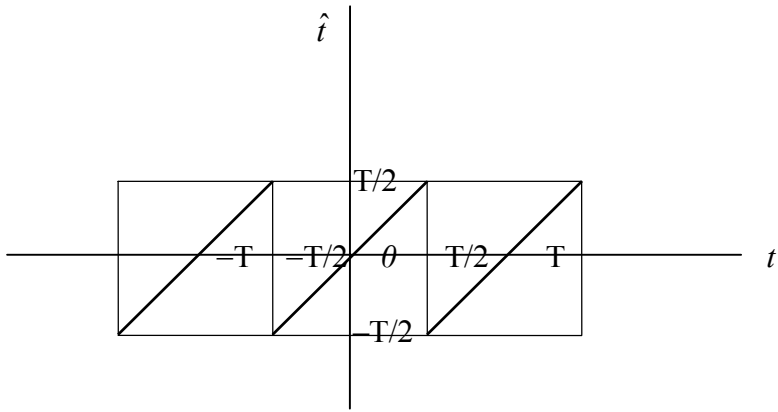

**Fig.2.** *The periodic saw-tooth function for the time operator* in the case of eq.(26).

This choice is convenient because the periodic function (26) for the time operator is a linear (increasing) function of time $t$ within *each* Poincaré interval; i.e., time flows always forward and preserves its usual meaning of *order* parameter for the system evolution.

The commutation relation of the energy and time operators, both *selfadjoint,* acquires in this case (discrete energies and periodic functions of $t$) the form:

$$[\hat{E},\hat{t}]= i\hbar\{1- T \sum_{n=0}^{\infty} \delta (t-[2n+1]T )\}. \tag{27}$$

Let us recall (cf., e.g. ref. [32]) that a generalized form of the uncertainty relation

$$(\Delta A)^2 \cdot (\Delta B)^2 \geq \hbar^2 [ <N> ]^2 \tag{28}$$

holds for two selfadjoint operators $\hat{A}$ and $\hat{B}$, which be canonically conjugate to each other through the more general commutator

$$[\hat{A},\hat{B}]= i\hbar \hat{N}, \tag{29}$$

$\hat{N}$ being a *third* selfadjoint operator. Then, from eq.(27) one can easily obtain that

$$(\Delta E)^2 \cdot (\Delta t)^2 \geq \hbar^2 \left[ 1 - \frac{T|\psi(T/2+\gamma)|^2}{\int_{-T/2}^{+T/2}|\psi(t)|^2 \, dt} \right], \tag{30}$$

where the parameter γ (it being a constant between –T/2 and +T/2) is introduced, in order to get a single-valued integral in the r.h.s. of eq.(27), running over *t* between –T/2 and +T/2: cf. refs.[33,35].

From eq.(30) it follows that, when $\Delta E \to 0$ (i.e. when $|g_n| \to \delta_{nn'}$), the r.h.s. of eq.(30) tends to zero since $|\psi(t)|^2$ tends to a constant value. In this case the distribution of the instants of time at which the wave-packet passes through point *x* becomes uniform (flat) within each Poincaré cycle. When $\Delta E >> D$ and $|\psi(T+\gamma)|^2 << (\int_{-T/2}^{T/2} |\psi(t)|^2 \, dt)/T$, the periodicity condition may become inessential whenever $\Delta t << T$; in other words, the more general uncertainty relation (30) transforms into the ordinary uncertainty relation (7) for systems with continuous spectra.

In the energy representation, the expression for the time operator (26) becomes a little bulky. But it is worthwhile to write it down, also for theoretical reasons. If one evaluates the mean value <t(x)> of the time instant at which the wavepacket passes through point *x*, then a long series of algebraic calculations leads to the expression [6]

$$\hat{t} = \frac{i\hbar}{2} \sum_{n;>n} (-1)^{N_n - N_{n'}} \frac{\vec{\Delta}_{n'}}{\Delta_{n'} \varepsilon_n} \tag{31}$$

where $N_n = (E_n - E_0)/D$, and where bilinear operators once more appear. In particular, now, the (finite-difference) operator $\vec{\Delta}_n$ means

$$A_n^* \vec{\Delta}_{n'} A_n \equiv A_n^* \Delta_{n'} A_n - A_n \Delta_{n'} A_n^*, \qquad \Delta_{n'} A_n \equiv A_{n'} - A_n \ .$$

Eventually, one obtains

$$<t(x)> = \sum_{n=0}^{\infty} g_n^* \varphi_n^*(x) \, \hat{t} \, g_n \varphi_n(x) \, / \sum_{n=0}^{\infty} |g_n \varphi_n(x)|^2.$$

Operator (31), in the simple case of two levels (*n*=0,1), acquires the simpler form

$$\hat{t} = \frac{-i\hbar}{2} \frac{\vec{\Delta}}{\Delta\varepsilon} , \qquad (31a)$$

while, when $D \equiv E_1 - E_0 \to 0,$ expression (31a) transforms into the differential form

$$\hat{t} = \frac{-i\hbar}{2} \frac{\vec{\partial}}{\partial\varepsilon} \qquad (31b)$$

which is quite similar to eq.(1b), which was found (for the first time in ref.[3]) for continuous energy spectra.

In all *realistic cases*, however, such as for all excited states of nuclei, atoms, molecules, etc., the energy levels are *not* regularly spaced, and moreover the levels themselves are not strictly defined because they do not correspond to discrete levels, but rather to resonances (a circumstance that happens *very* frequently, at least as a consequence of photon decays or emissions): so that not even the duration of the Poincarè cycle is exactly defined. When the resonances are large, one practically goes back to the continuous case. By contrast, in the case of very narrow resonances, when the level widths $\Gamma_n$ are much smaller than the level spacings $|E_n - E_{n'}|$, which again corresponds to many realistic systems —such as nuclei, atoms and molecules in their low-energy excitation regimes—, we can introduce an approximate description (with any desired degree of accuracy, even of the order $\Gamma_n / |E_n - E_{n'}|$) in terms of quasi-cycles with a quasi-periodic evolution: And, for sufficiently long time intervals, the motion inside such systems can be regarded, with the same accuracy, as *a periodic motion*. Then, *quasi-selfadjoint time operators* of the type (26) or (31) can be introduced, and all the relevant temporal quantities defined and calculated (always within an accuracy of the order $\Gamma_n / |E_n - E_{n'}|$).

Let us observe that, if a system has a partially continuous and a partially discrete energy-spectrum, one can easily use expressions (1) for the continuous energy spectrum, and expressions (26) and (31) for the discrete energy spectrum.

### 8. Conclusions, perspectives, and applications.

1. Time *t*, as well as space *x*, is known to play sometimes the role of a parameter, while in other cases they represent quantities which one wishes to measure, and therefore must correspond in quantum mechanics to operators. For instance, we actually regard time *t* as an observable when we have to measure flight-times, collision durations, tunnelling times, interaction-durations, mean life-times of metastable states, and so on (see, e.g., refs.[5,16,17,38] and refs. therein).

The (maximal) hermitian Time Operator (1b) discussed in this paper does possess a general validity, for any quantum collision or motion processes in the *continuum* range of the energy spectrum, and both in non-relativistic quantum mechanics and in one-dimensional quantum electrodynamics. It cannot be defined (unless one goes on to bilinear operators) in the cases with zero fluxes or with particles at rest: but for those cases there are no evolution processes at all, so that the above condition does not really imply any loss of generality. Moreover, the *uniqueness* of the time operator (1b) does directly follow from the uniqueness of the Fourier-transformation linking the time with the energy representation.

---

[6] Of course, one has to average over the flux density, but for simplicity in this case it is possible to average over $|\Psi(x,t)|^2$.

As we were saying, operator (1b) *has been* already rather fruitfully when *applied* for defining and evaluating the Tunneling Times (cf. refs [16,17, 31]), as well as for the time analysis of nuclear reactions (see, for instance, refs [5,38], and the first one of refs.[6]).

2. In the *discrete* range of the energy spectrum, the Time Operator assumes the form (31) in the energy representation, and the form (26) in the time representation (when it acts on the space of the wavepackets representing superpositions of bound states: in full analogy with the situation for the azimuth-angle operator). Such a Time Operator cannot be defined, however, in the case of just *one* bound-state, since also in this case there is no evolution.

When one deals with overlapping resonances or with infinitesimally close levels, formula (31a) transforms into formula (1b), exploited above for systems with continuous spectra.

3. The commutation relations (4) and (21), and also the uncertainty relations (7) and (30), play exactly the same role of the analogous relations known to exist for other pairs of canonically conjugate observables (such as coordinate $\hat{x}$ and momentum $\hat{p}_x$, in the case of eq.(7); and as azimuth angle $\varphi$ and angular momentum $\hat{L}_z$, in the case of eq.(30)). Incidentally, relations (21) and (30) do not replace, but rather *extend* (and include) the time and energy uncertainties given by Krylov and Fock[36]. Moreover, they are consistent with the conclusions by Aharonov and Bohm[37]. Our formalism can help attenuating the endless debates about the status of the time-energy uncertainty relation.

4. Finally, let us observe that *not only the time operator, but any other quantities corresponding to (maximal) hermitian operators* (like momentum in a semi-space with a rigid wall, and like the radial momentum in free space, both defined over a semi-bounded axis going from 0 to ∞ only) can be regarded as *quantum observables,* in the same way as the quantities to which selfadjoint operators correspond: *without the need of introducing any new physical postulates.* A fortiori, the same conclusion is valid for the quasi-selfadjoint operators, like (26) and (31), met for quasi-discrete spectra (like for regions of very narrow resonances).

**9. Acknowledgements.** The authors thank L.Fraietta, A.S.Holevo, V.L.Lyuboshitz, V.Petrillo, G.Salesi, E.Spedicato, M.T.Vasconcelos, B.N.Zakhariev and M.Zamboni-Rached for discussions or kind collaboration.

# APPENDIX A

### Another example of a non-selfadjoint operator which is a quantum observable

After that we have seen the good properties of our Time operator (1b), one might ask himself why a time operator was not introduced in standard quantum mechanics, even if quantum mechanics is known to associate an operator to every observable. The reason, as we have seen, is that operator (1b) is defined as acting on the space *P* of the continuous, differentiable, square-integrable functions *f* that satisfy the conditions

$$\int_0^\infty |f|^2 \, dE < \infty; \quad \int_0^\infty |\partial f / \partial E|^2 < \infty; \quad \int_0^\infty |f|^2 E^2 \, dE < \infty,$$

which is a space dense[28] in the Hilbert space of $L^2$ functions defined (only) over the interval $0 \leq E < \infty$. Therefore, operator (1b) is not selfadjoint for the fact that $P$ is *not* the space of the fuctions of $E$ defined over the whole $E$-axis; as a consequence, operator (1b) does not allow an (orthogonal) identity resolution. Essentially because of these reasons, Pauli[1] rejected the use of a Time operator: and this had the effect of practically stopping studies on this subject for about forty years.

However, as already mentioned in the text, von Neumann[28] had claimed that considering in quantum mechanics only selfadjoint operators could be too restrictive. To clarify this issue, let us quote an explanatory example set forth by von Neumann himself. Let us consider a particle Q, free to move in a semi-space bounded by a rigid wall (see Fig.1). We shall then have $0 \leq x < \infty$. Consequently, the impulse $x$-component of particle Q, which reads

$$\hat{p}_x = -i\hbar \frac{\partial}{\partial x} \; ,$$

will be a non-selfadjoint but only a (maximal) hermitian operator: nevertheless, it is an observable with an obvious physical meaning. All the same can repeated for our Time operator (1b).

# APPENDIX B

## On the bilinear Time operator

With the aim of making quantum mechanics as 'realistic' as possible, one may adopt a *space-time description* of the collision phenomena, by introducing wave packets. As soon as a space-time descriptions of interactions has been accepted, one can immediately realize that, even in the framework of the usual wave-packet formalism, a quantum operator for the observable time is operating (as it was firstly noticed in ref.[22]). Namely, it was implicitly used for calculating the packet time-coordinate, the flight-times, the interaction-durations, the mean-lifetimes of metastable states, the tunnelling times, and so on (see refs.[2,17,38], as well as the first one of ref.[22], and the last one of refs.[16]). A preliminary, heuristic inspection of the adoption of the formalism suggests the adoption of the following operators[2]

$$\hat{t}_1 = -i\hbar \frac{\partial}{\partial E}; \quad \hat{t}_2 = (-i\hbar/2) \frac{\overset{\leftrightarrow}{\partial}}{\partial E} \quad [E \equiv E_{\text{tot}}] \tag{B.1}$$

acting on a wave-packet space which we must carefully define [because of the differential character of these different forms of the Time Operator]. We are going to discuss this point.

Let us consider, for simplicity, a *free* particle in the *one-dimensional* case, i.e. the packet

$$F(t,x) = \int_0^\infty dk \cdot \tilde{F}(E,k) \cdot \exp[i(kx - Et/\hbar)] \tag{B.2}$$

where $E \equiv k^2\hbar^2/m_0$. The integral runs only over the positive values owing to the 'boundary' conditions imposed by the initial (*source*) and final (*detector*) experimental devices. Notice that, in so doing, we chose as the frame of reference that one in which source and detectoe are at rest: i.e. the laboratory. In particular, let us consider for simplicity the case of source and detector *at rest one with respect to the other*.

One can observe that the packet (average) position is always to be calculated at a fixed time $t = \bar{t}$; analogously, the wave packet time-coordinate is always to be calculated (by suitably averaging over the packet) for a position $x = \bar{x}$ along a particular packet-propagation-ray. Therefore, in our case we can fix a particular $x = \bar{x}$, and restrict ourselves to considering, instead of the packets (B.2), the functions

$$F(t, \bar{x}) = \int_0^\infty dp\, f'(p, \bar{x}) \exp(-iEt/\hbar) = \int_{(+)} dp\, f(E, \bar{x}) \exp(-iEt/\hbar), \quad (B.3)$$

where $E \equiv E_{tot} \equiv E_{kin} \equiv p^2/m_0$; $p \equiv k\hbar$; and $f' = f\, dE/d|p|$. Quantities $F(t, \bar{x})$ and $f(E, \bar{x})$, being *only* functions either of $t$ or of $E$, respectively, are neither wave functions (that satisfy any Schroedinger equation), nor do they represent states in the chronotopic or 4-momentum spaces. Let us briefly set

$$F \equiv F(t) \equiv F(t, \bar{x}); \quad f \equiv f(E) \equiv f(E, \bar{x}). \quad (B.4a)$$

It is easy to go from functions *F*, or *f*, back to the 'physical' wave packets, so that one gets a one-to-one correspondence between our functions and the 'physical states'. We shall respectively call «space *t*» and «space *E*» the functional spaces of the *F*'s and of the transformed functions *f*'s, with the mathematical conditions that we are going to specify. In those spaces, for example, the norms will be

$$\|F\| \equiv \int_{-\infty}^{\infty} |F|^2\, dt; \quad \|f\| \equiv \int_0^\infty |f|^2\, dE. \quad (B.4b)$$

In any case, due to equations (B.3), the space *t* and the space *E* are representations of the same abstract space *P*

$$F \to |F>; \quad f \to |f>, \quad (B.4c)$$

were $|F> \equiv |f>$. Let us now *specify* what has been previously said by assuming space *P* to be the space of the continuous, differentiable, square-integrable functions $f$ that satisfy the conditions:

$$\int_0^\infty |f|^2\, dE < \infty; \quad \int_0^\infty |\partial f/\partial E|^2 < \infty; \quad \int_0^\infty |f|^2 E^2\, dE < \infty, \quad (B.5)$$

a space that we know to be *dense*[28] in the Hilbert space of $L^2$ functions defined over the interval $0 \leq E < \infty$.

Still within the framework of ordinary quantum mechanics dealing with wave packets, let us define in the most natural way

$$<t(\bar{x})> \equiv \frac{\int_{-\infty}^{\infty} j(x,t) t\, dt}{\int_{-\infty}^{\infty} j(x,t)\, dt} = \frac{\int_0^\infty dE \frac{1}{2}[F^*(x,E)(\hat{t}v + v\hat{t})F(x,E)]}{\int_0^\infty dE v |F(x,E)|^2}, \quad (B.6)$$

when going on from the time to the energy representation. Quantity $v$ is the velocity $p/m_0$. In equation (B.6), quantity $F(x,E)$ is the Fourier-transform of the moving 1-D wave packet

$$\Psi(x,t) = \int_0^\infty F(x,E) \exp(-iEt/\hbar)\, dE = \int_0^\infty f(E)\, \varphi(x,E) \exp(-iEt/\hbar)\, dE \tag{B.7}$$

with the normalization condition

$$\int_0^\infty v|G(x,E)|^2\, dE = \int_0^\infty v|g(E)|^2\, dE = 1 ,$$

while the flux density

$$j(x,t) = \mathrm{Re}\, [\Psi^*(x,t)\, (\hbar/i\mu)\, \partial\Psi(x,t)/\partial x] \tag{B.8}$$

refers to the wavepacket (B.7). In the particular case of free motions, in eq.(B.7) one has: $F(x,E) = f(E)\exp(ikx)$; $\varphi(x,E) = \exp(ikx)$; and $E = \mu\hbar^2 k^2/2 = \mu v^2/2$.

One can easily verify, by direct calculations, that eq.(B.6) implies as time operator the expression $\hat{t} = (-i\hbar/2)\dfrac{\overleftrightarrow{\partial}}{\partial E}$; in other words, this suggests adopting as the Time Operator the bilinear *derivation*

$$\hat{t} \equiv \hat{t}_2 \equiv (-i\hbar/2)\frac{\overleftrightarrow{\partial}}{\partial E}\ . \tag{B.9}$$

By easy calculations, one realizes that one can also adopt the (standard) operator

$$\hat{t} \equiv \hat{t}_1 \equiv -i\hbar\frac{\partial}{\partial E}\ , \tag{B.10}$$

but at the price of imposing on space $P$ the *subsidiary condition* $f(0, \bar{x}) = 0$.

In order to use in details the bilinear derivation (B.9) as a (bilinear) operator, it would be desirable to introduce a careful, new formalism. Here, however, we limit ourselves for brevity's sake at referring to refs. [39], were also the case of a space-time, four-position operator (besides of the 3-position operator) was exploited.

# *R e f e r e n c e s*